# Defects of Ge quantum dot arrays on the Si(001) surface

## V.A. Yuryev,[*] L.V. Arapkina

*A.M. Prokhorov General Physics Institute of RAS, Moscow, 119991, Russia*

**Abstract**

Defects of Ge quantum dot arrays may affect the electrophysical, photoelectrical or optical properties of Ge/Si heterostructures as well as the functionality of devices produced on their basis. The defects of Ge quantum dot arrays formed at moderate temperatures on the Si(001) surface have been investigated by the ultra high vacuum scanning tunnelling microscope integrated with the molecular beam epitaxy chamber. A preliminary classification of the defects has been carried out. Morphological peculiarities of the defects have been studied. The surface densities of defects of different types have been determined.

## 1. Introduction

Dense arrays of Ge quantum dots (QD) grown on the Si(001) surface at moderate temperatures attract an attention of researchers for the past decade [1, 2] due to their great potential in microelectronics and photonics [3]. Recent technological achievements enabled the controllable formation of the 2-D QD arrays with the required Ge cluster densities [4, 5]. However, three highly important issues are still far from solution—the uniformity of cluster types in the arrays, the dispersion of cluster sizes and the array defectiveness. This article is devoted to the latter problem.

Ge clusters with parameters considerably different from the characteristic parameters of clusters in the array or areas with appreciable local deviation of array parameters from those typical for the conditions of the array formation will hereinafter be referred to as defects of Ge quantum dot arrays. The defects of Ge QD arrays may affect the electrophysical, photoelectrical or optical properties of Ge/Si heterostructures as well as the functionality of devices produced on their basis. A large Ge cluster may not only violate the homogeneity of an array but, that is much worse, it can pierce a spacer between layers of a multilayer structure causing shortcut of the QD layers and, e.g., detector leakage or, if the spacer is thick enough to keep integrity, give rise to so high local strain in the Si spacer that disturb the growth of the next Ge QD layers. Remark, that this may be a crucial issue for the ordered QD arrays [6] taking into account the requirements of uniformity they must meet to form 3-D crystals of artificial atoms with precisely reproduced regularity in consecutive QD layers carefully separated by rather thin Si spacers.

## 2. Experimental

The experiments were carried out using an ultra high vacuum instrument consisting of the Riber EVA 32 molecular beam epitaxy (MBE) chamber coupled with the GPI-300 high resolution scanning tunneling microscope (STM) which enables *in situ* sample study at any stage of processing serially investigating the surface structure down to atomic level and giving additional treatments to the specimen.

The surfaces of B-doped Cz Si (100) substrates ($\rho = 12\ \Omega\ cm$) were completely purified of the oxide film as a result of short annealing at the temperature of $\sim 925°C$ [7]. Germanium was deposited directly on the deoxidized silicon surface from the crucible with the electron beam evaporation.

The rate of Ge deposition was $\sim 0.15$ Å/s; the Ge coverage ($h_{Ge}$) was varied from 6 to 14 Å for different samples. The substrate temperature ($T_{gr}$) was 360 or 530°C during the process. The rate of the sample cooling down to the room temperature was $\sim 0.4°C/s$ after the deposition. Images were scanned at room temperature in the constant tunneling current ($I_t$) mode. The STM tip was zero-biased while a sample was positively or negatively biased ($U_s$) for filled or empty state imaging. The details of the sample preparation as well as the experimental techniques can be found elsewhere [7]. The WSxM software [8] was used for STM image processing.

---

[*] Corresponding author. Tel.: +7-499-503-8144; fax: +7-499-135-0356; e-mail: vyuryev@kapella.gpi.ru

## 3. Results and discussion

As a result of the STM study of the samples grown at different values of $h_{Ge}$ and $T_{gr}$, a preliminary classification of the defects was carried out. The revealed defects were subdivided into the following groups: (i) large hut or dome clusters [1, 9, 10] with the sizes untypical for the growth temperature referred to as *large defects* (Fig. 1), (ii) extended clusters stretching along the <110> direction (*extended defects*), which are shown in Fig. 2(a), and (iii) areas in which QDs did not grow (*depleted regions*) demonstrated in Fig. 2(b).

Fig. 1(a) presents a single large defect surrounded by the regular hut clusters. Large defects often gather in groups like those shown in Fig. 1(b). Another large defect looking like some "rounded" cluster is depicted in Fig. 1(c). Such defects being parts of arrays strongly violate their homogeneity. It should be noted also that the large defects never cause depletion areas in the QD arrays formed at $T_{gr} = 360°C$. In the arrays grown at $T_{gr} = 530°C$, some depletion was observed around the large defects and their groups.

Fig. 2(a) demonstrates a picturesque view on a huge extended defect. Heights of these defects $h > 20$ nm, lengths $l > 300$ nm, width $w \sim 100–200$ нм. The hut cluster density is seen to dramatically drop in the vicinity of the extended defect. It is a common rule: the extended defects deplete the adjacent areas consuming all Ge atoms arriving onto the surface near them. Therefore, the extended defects may affect the QD array parameters and at high enough density sufficiently decrease the QD density throughout the array. The extended defects formation probability is increased as $T_{gr}$ is enhanced.

Morphological peculiarities of the large and extended defects have been investigated. Fig. 3(a) depicts a profile of cross section of the extended defect shown in Fig. 2(a). It should be noted that these defects are observed to have very sophisticated shape and faceting. Fig. 3(a) presents only three facets: {102} tilted $\sim 26°$ to the (001) plane [9], {106} sloped $\sim 9.5°$ and an exotic {3 0 10} face inclined $\sim 16.5°$ to (001). Other profiles revealed {105} and {113} facets as well as some additional exotic highly inclined facets. Although faceting of extended defects to some extent resembles faceting of the dome clusters [9, 10] their nature is absolutely different from the nature of ordinary domes.

The STM image of a facet of the large defect shown in Fig. 1(c) is presented in Fig. 3(b); steps and (001) terraces are resolved. The cluster looking like "rounded" have appeared to be faceted. The facet empty state image resembles that of the {105} one rather than some other high indexed facet characteristic for the dome clusters [11].

Facets of the extended defects have been investigated with atomic resolution. The periodical structure of the dimer pairs on the (001) terraces are resolved in the STM image of a facet of the extended defect given in Fig. 3(c). The period was measured to be $\sim 11$ Å along the terraces and $\sim 7$ Å in the normal direction (from apex to base).

In some samples, areas in which the Ge clusters did not grow were revealed. We did not manage to obtain the STM images of the surface in such depleted regions. These defects are likely strongly contaminated areas which are failed to be refined by the standard procedure of the Si surface pre-growth cleaning. Fig. 2(b) demonstrates the STM image of the sample containing the described type of defects.

Fig. 4 depicts a plot of the defect surface density for different values of $h_{Ge}$ and $T_{gr}$. At different $T_{gr}$, the density of the large defects is seen to differently change with $h_{Ge}$. At $T_{gr} = 530°C$ their density is approximately constant ($\sim 2 \times 10^9$ cm$^{-2}$) whereas at $T_{gr} = 360°C$ it is maximum at $h_{Ge} = 8$ Å ($\sim 4 \times 10^9$ cm$^{-2}$), drastically drops at $h_{Ge} = 10$ Å ($\sim 4 \times 10^7$ cm$^{-2}$) and increases again at $h_{Ge} = 14$ Å ($\sim 10^9$ cm$^{-2}$). The density of extended defect was evaluated as $10^6$ cm$^{-2}$.

The origin of the defects has not been found out thus far. We suppose only that extended defects originate from some extended structural defects of the initial Si (100) substrates, such as stocking faults (SF). An observation was made that supports this opinion: The extended defects have mainly been observed in the sample subjected to a long term annealing at $T = 500–600°C$ before Ge deposition. It is known that high temperature treatments, like that applied for the surface deoxidization, may give rise to the SF nuclei. Then in the process of cooling and further annealing SF grow and ripen reaching appropriate sizes to become the seeds on which the extended defects nucleate.

**Figure captions**

Fig. 1. STM images of the large defects ($h_{Ge}$ = 14 Å; $T_{gr}$ = 360ºC): single (a) and grouped (b) large defects, a "rounded" large defect (c).

Fig. 2. STM images of the extended defect (a) ($h_{Ge}$ = 11 Å, $T_{gr}$ = 530ºC) and the depleted regions (b) ($h_{Ge}$ = 8 Å, $T_{gr}$ = 360ºC); the large defects are also seen in the image (b) as tailed light spots.

Fig. 3. Morphology of the large and extended defects: a cross section profile (a) of the extended defect shown in Fig. 2(a), facets are emphasized by solid lines (tilt to the (001) plane is ~ 26° for the left side and ~ 9.5° and ~ 16.5° for the right side); STM image of a facet of the large defect (b) shown in Fig. 1(c), steps and (001) terraces are resolved; a fine structure of the (001) terraces of the extended defect facet (c), the terraces are running from the lower left to the upper right corner of the image, dimer pairs are resolved.

Fig. 4. Surface density of large (▲, $T_{gr}$ = 530ºC; ●, $T_{gr}$ = 360ºC) and extended (■, $T_{gr}$ = 530ºC) defects estimated from STM images.

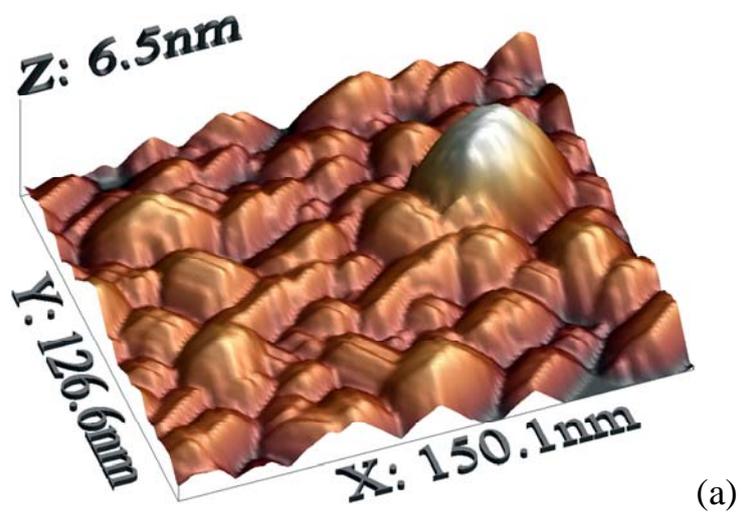

(a)

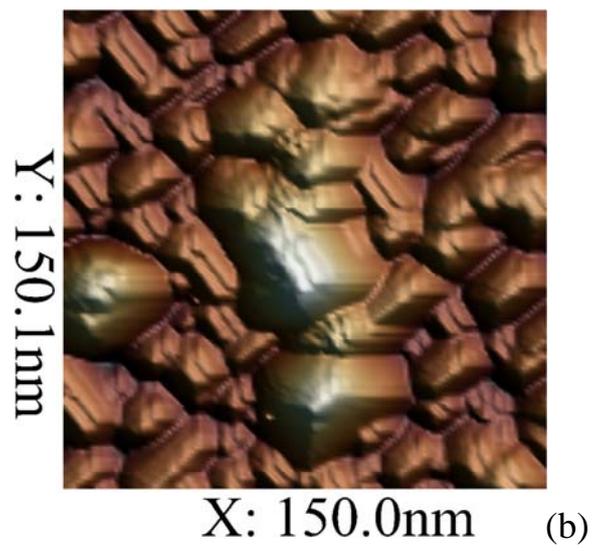

(b)

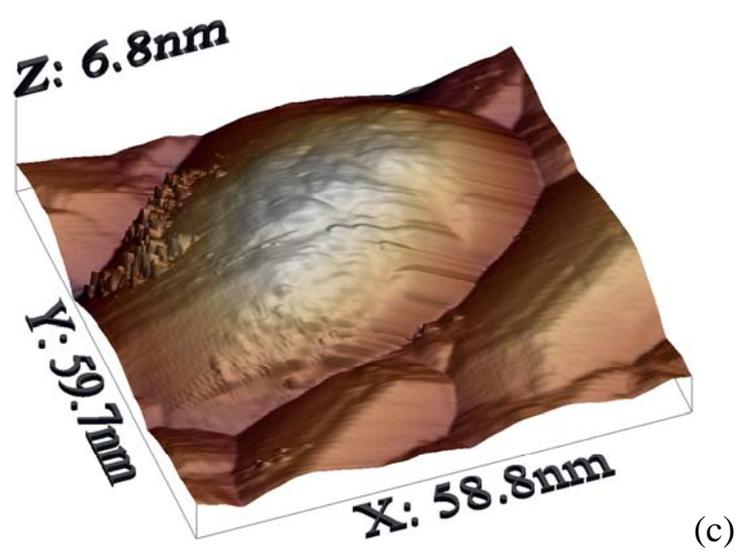

(c)

Figure 1.

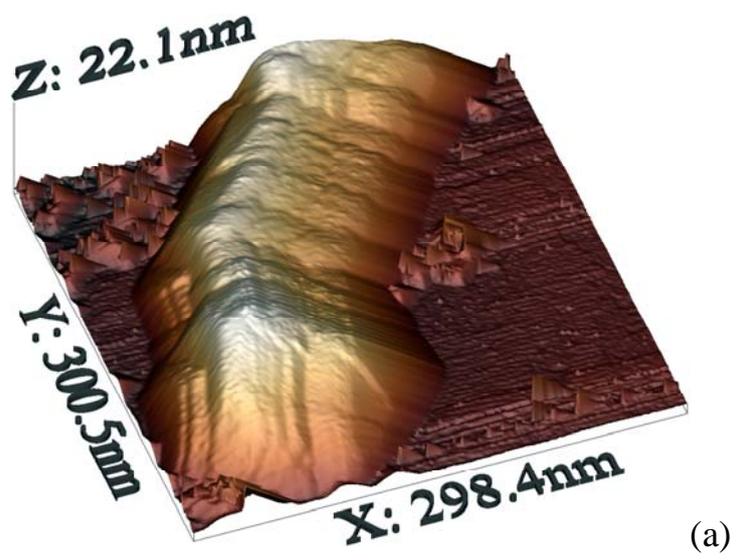

(a)

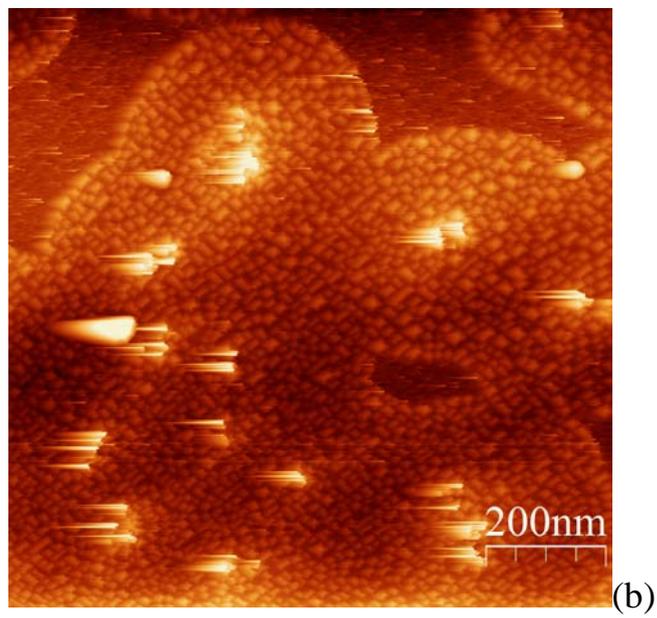

200nm

(b)

Figure 2.

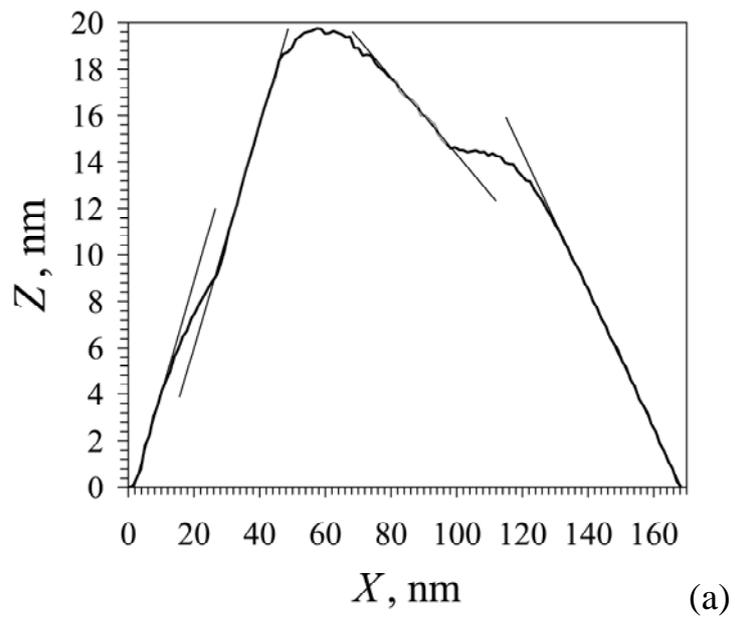

(a)

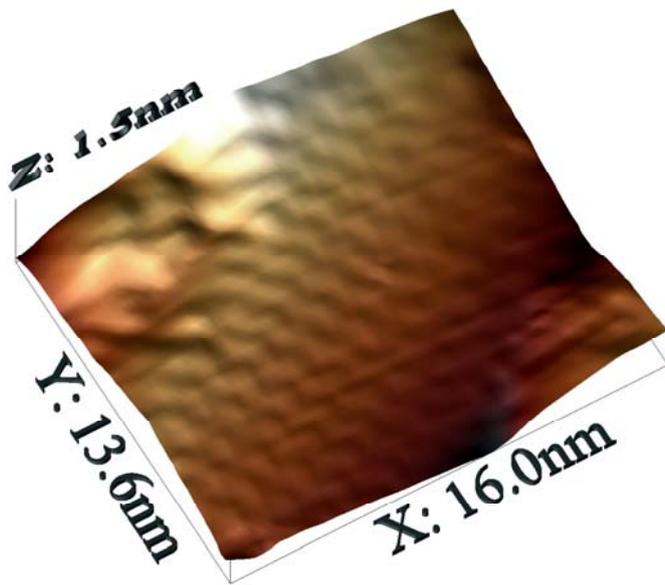

(b)

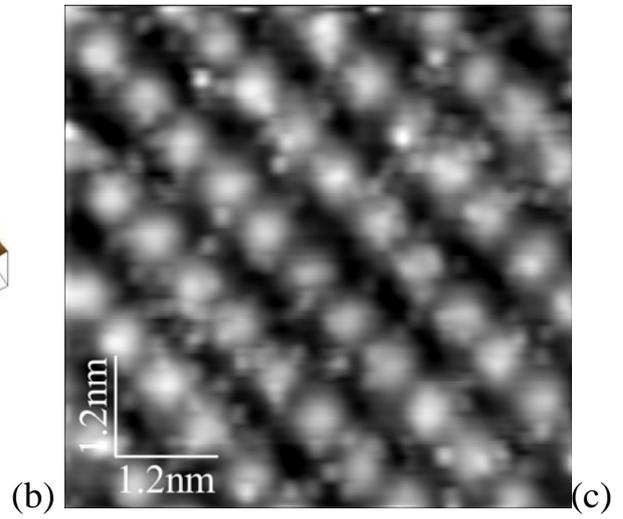

(c)

Figure 3.

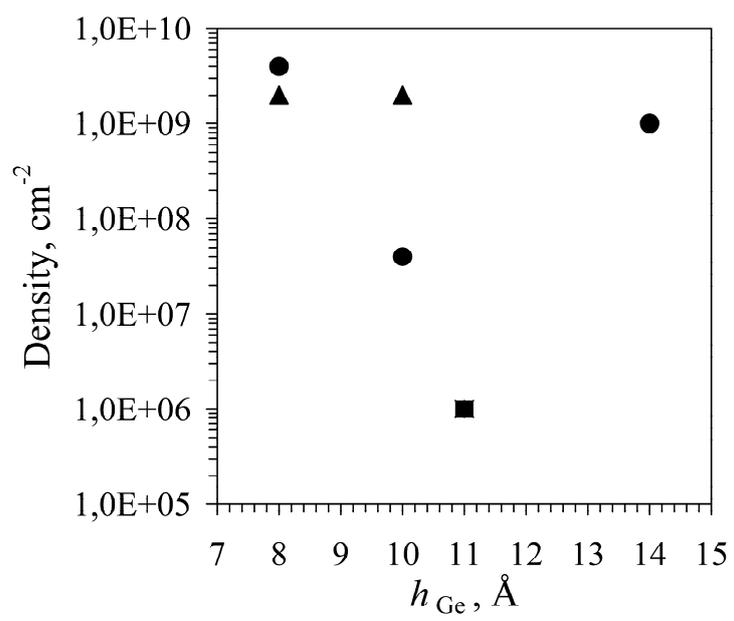

Figure 4.